# Superconductivity in pressurized CeRhGe$_3$ and related non-centrosymmetric compounds

Honghong Wang[1,6]*, Jing Guo[1]*, Eric D. Bauer[2], Vladimir A Sidorov[3], Hengcan Zhao[1], Jiahao Zhang[1], Yazhou Zhou[1], Zhe Wang[1], Shu Cai[1], Ke Yang[4], Aiguo Li[4], Xiaodong Li[5], Yanchun Li[5], Peijie Sun[1], Yi-feng Yang[1,6,7], Qi Wu[1], Tao Xiang[1,6,7], J. D. Thompson[2]†, Liling Sun[1,6,7]†

[1]*Institute of Physics, Chinese Academy of Sciences, Beijing 100190, China*
[2]*Los Alamos National Laboratory, MS K764, Los Alamos, NM 87545, USA*
[3]*Institute for High Pressure Physics, Russian Academy of Sciences, 142190 Troitsk, Moscow, Russia*
[4]*Shanghai Synchrotron Radiation Facilities, Shanghai Institute of Applied Physics, Chinese Academy of Sciences, Shanghai 201204, China*
[5]*Institute of High Energy Physics, Chinese Academy of Science, Beijing 100049, China*
[6]*University of Chinese Academy of Sciences, Beijing 100190, China*
[7]*Collaborative Innovation Center of Quantum Matter, Beijing, 100190, China*

We report the discovery of superconductivity in pressurized CeRhGe$_3$, until now the only remaining non-superconducting member of the isostructural family of non-centrosymmetric heavy-fermion compounds CeTX$_3$ (T = Co, Rh, Ir and X = Si, Ge). Superconductivity appears in CeRhGe$_3$ at a pressure of 19.6 GPa and the transition temperature $T_c$ reaches a maximum value of 1.3 K at 21.5 GPa. This finding provides an opportunity to establish systematic correlations between superconductivity and materials properties within this family. Though ambient-pressure unit-cell volumes and critical pressures for superconductivity vary substantially across the series, all family members reach a maximum $T_c^{max}$ at a common (±1.7%) critical cell volume $V_{crit}$, and $T_c^{max}$ at $V_{crit}$ increases with increasing spin-orbit coupling strength of the *d*-electrons. These correlations show that substantial Kondo hybridization and spin-orbit coupling favor superconductivity in this family, the latter reflecting the role of broken centro-symmetry.

The discovery of the first heavy-fermion superconductor $CeCu_2Si_2$ in 1979 [1] opened the new field of unconventional superconductivity and posed a fundamental challenge for understanding the superconducting mechanism. After nearly forty years, more than thirty heavy-fermion superconductors have been found, and there is growing appreciation that their superconductivity emerges at the border of competing or coexisting electronic orders [2], suggesting that fluctuations of these orders may be the source of superconductivity. Among heavy-fermion compounds, the $CeTX_3$ (T = Co, Ir, Rh and X = Si, Ge) [3,4] family belongs to a subset whose crystallographic structure (*I4mm*) lacks a center of inversion symmetry [5,6]. Except for the mixed-valence member $CeCoSi_3$, others of the family order antiferromagnetically at atmospheric pressure. Application of pressure to $CeRhSi_3$, $CeIrSi_3$, $CeCoGe_3$ and $CeIrGe_3$ suppresses their Néel temperature toward zero temperature where a dome of superconductivity emerges [7-18]. These observations have led to suggestions that magnetic fluctuations may induce unconventional superconductivity, but this remains unsettled, in part because of the possible non-trivial role of antisymmetric spin-orbit coupling that is a consequence of these materials' non-centrosymmetry [6]. Beside its possible role in forming superconductivity, sufficiently strong spin-orbit coupling would also lead to an unusual pairing state that is a mixture of spin-singlet and spin-triplet components.

As a member of the magnetic CeTX$_3$ family, CeRhGe$_3$ is expected to become superconducting under pressure. However, experiments to about 8 GPa find only a monotonic increase in its Néel temperature [15]. Here, we report the first observation of pressure-induced superconductivity in CeRhGe$_3$. As will be discussed, this discovery yields a broader perspective on conditions favoring superconductivity in this family of nominally isoelectronic compounds. Details of crystal growth and high-pressure procedures are given in the Supplementary Information.

Specific heat and magnetic susceptibility measurements on polycrystalline CeRhGe$_3$ reveal three magnetic transitions at 14.6, 10 and 0.55 K, although the transition at the lowest temperature is not clearly discernible in electrical resistivity [4]. In the present study, the first two transitions at $T_{N1}$ and $T_{N2}$ are identified from minima in second derivatives of the resistivity (Fig.S1a of the Supplementary Information) and as kinks in resistance versus temperature that is plotted in Fig. 1 for pressures to 26.5 GPa. With increasing pressure, $T_{N1}$ and $T_{N2}$ move to higher temperatures, consistent with the previously reported pressure dependence of $T_{N1}$ [15]. $T_{N1}$ reaches a maximum at 8.5 GPa and $T_{N2}$ is maximized near 10.5 GPa. Both transitions decrease at higher pressures and merge at ~13.7 GPa within experimental resolution. This is seen most clearly in the pressure evolution of $\partial^2 R/\partial T^2$ shown in Fig. S1a. The merged AFM transition temperature $T_N$ falls monotonically with increasing pressure up to 18.6 GPa (Fig.1b and Fig.S1b of the Supplementary Information). This merging of two antiferromagnetic transitions and its pressure dependence are similar to that reported for CeIrGe$_3$ [12].

At 19.6 GPa, where $T_N$ = 3.6 K (as indicated by the arrow in Fig. 1c), there is a pronounced drop in resistance at ~1.1 K. This resistance drop is largest at 21.5 GPa, where the resistance declines by ~76.6 %, and then the drop becomes smaller upon further compression. This observation is confirmed in a separate measurement in which a pressure-induced resistance drop is ~92.8 % at 22.0 GPa (inset of Fig. 1c). To explore the origin of this resistance drop, we applied a magnetic field to CeRhGe$_3$ subjected to 21.5 GPa and cooled the high-pressure cell to 40 mK. As shown in Fig. 1d, the onset temperature of the resistance drop shifts to lower temperature upon increasing magnetic field, suggesting that this pressure-induced resistance drop in CeRhGe$_3$ results from a superconducting transition. As shown in Fig.S2 of Supplementary Information, the rate of decrease in the onset temperature is very high (-11.6 T/K), which is a common feature of the remarkably weak temperature dependence of the upper critical field in other CeTX$_3$ compounds [5]. Furthermore, the *ac* susceptibility measured at 20.1 GPa (the inset of Fig.1d) becomes diamagnetic at 1.2 K. All these measurements show that the pressure-induced resistance drop in low temperatures results from a superconducting transition. The resistance below the onset temperature is still finite due to the presence of sample micro-cracks generated by the quasi-hydrostatic pressure environment (See Supplementary Information for details).

This pressure-induced superconductivity emerges in the non-centrosymmetric *I4mm* crystal structure. Results of synchrotron X-ray diffraction (XRD) measurements on CeRhGe$_3$ at pressures up to 28.5 GPa are discussed in the Supplementary

Information and summarized in Fig. 2. With increasing pressure, the lattice constants decrease, but the crystal structure remains unchanged. A fit of the pressure-dependent cell volume to a Murnaghan equation of state ($V = V_0 \times [1 + P \times (B'/B)]^{-1/B'}$) gives a bulk modulus ($B$) and its pressure derivative ($B'$) of 115 GPa and 5, respectively. Here $V$ and $V_0$ are the high-pressure and zero-pressure volumes of the material, respectively. The stability of the structure against pressure rules out the possibility that the superconductivity emerges in a different crystal structure at low temperatures. This is consistent with the observation of the huge derivative of the upper critical field, $\partial H_{c2}/\partial T$, of CeRhGe$_3$ which is a common feature of these non-centrosymmetric superconductors.

Figure 3 shows the pressure versus temperature phase diagram obtained from our measurements. The overall bell-shaped response of antiferromagnetism to pressure can be understood using the Doniach's model of competing Ruderman-Kittel-Kasuya-Yosida (RKKY) and Kondo interactions [19]. With increasing pressure, the Kondo hybridization becomes dominant, which suppresses the RKKY-mediated long-range order. However, $T_N$ does not go continuously to zero temperature. A similar pressure dependence of the AFM transition temperature is also found in CeIrGe$_3$ [15]. Nevertheless, the substantially suppressed Néel order around 20 GPa leaves the possibility that magnetic fluctuations could play an important role in inducing superconductivity that initially coexists with the AFM order.

In Fig. 4 we place our discovery of superconductivity in CeRhGe$_3$ in the context of early results on other members of the CeTX$_3$ family [4-13,15,18] and compare the pressure dependence of the superconducting transition temperature for the whole CeTX$_3$ family. We find that the critical pressure $P_{crit}$ at which pressure-induced superconductivity reaches a maximum $T_c$ is strongly correlated with the unit-cell volume at the ambient pressure (Fig.4a). For X=Si or Ge, the cell volume increases in the sequence T= Co, Rh, Ir, and for a given T, the cell volume is larger for X=Ge. CeCoSi$_3$, which has the smallest cell volume, has a mixed-valence 4*f*-configuration with strong hybridization between *f* and conduction (*c*) electrons and is not superconducting above 0.5 K [20]. In contrast, the Kondo hybridization in the large-cell compounds, CeIrGe$_3$ and CeRhGe$_3$, is much weaker [4]. As the unit cell expands, the critical pressure $P_{crit}$ increases by an order of magnitude from CeRhSi$_3$ to CeIrGe$_3$. This correlation implies that an optimally strong Kondo hybridization induced by compression is prerequisite for superconductivity in this family of non-centrosymmetric heavy-fermion compounds.

Assuming that our equation of state with the bulk modulus and its derivative for CeRhGe$_3$ provides a reasonable approximation for estimating the cell volume of other CeTX$_3$ members, we can estimate the critical cell volume $V_{crit}$ at $P_{crit}$. Fig, 4b shows the maximum superconducting transition temperature $T_c^{max}$ as a function of $V_{crit}$. Interestingly, all $T_c^{max}$ fall within a rather narrow range of $V_{crit}$ values that vary by only ±1.7%, as emphasized by the vertical dashed lines in Fig. 4a. It appears, then, that there is an approximate optimal cell volume for achieving a maximum $T_c$. Because all the family members are nominally isoelectronic, this correlation could be an indication of an optimal Kondo hybridization for a maximum $T_c$. Further, as shown by the dashed lines in Fig. 4a, $V_{crit}$ places the optimal superconductivity near the cross-over to a mixed-valence regime found in CeCoSi$_3$, suggesting y that critical valence fluctuations may also play certain role in producing superconductivity in this CeTX$_3$ family.

A more striking observation is that that $T_c^{max}$ increases systematically in CeTX$_3$ as the outmost electrons of T change from 3$d$ to 4$d$, andto 5$d$. For T=Rh and Ir, $T_c^{max}$ is similar irrespective of the X element. For the T element, the spin-orbit interaction varies as is proportional to $Z^4/n^3$, where $Z$ is the atomic number and $n$ is the principal quantum number [6]. Thus the spin-orbit coupling increases when T varies from Co (n=3) to Rh (n=4), and Ir (n=5). If we set the spin-orbit coupling to 1 for T=Co, we find the spin-orbit coupling is ~3 and ~14 times stronger for T=Rh and Ir, respectively. Clearly, $T_c^{max}$ increases with increasing spin-orbit coupling strength.

The role of antisymmetric spin-orbit coupling is to lift spin degeneracy of single-particle states, removing parity symmetry and creating spin-split electronic bands in which spins are polarized tangential to the electrons' momentum. In general, band splitting due to antisymmetric spin-orbit coupling is band-dependent, and an effective antisymmetric spin-orbit coupling is expected to depend on the extent of hybridization between $f$ and conduction electrons [21]. Though a correlation between $T_c^{max}$ and magnitude of the atomic spin-orbit coupling is obvious in Fig. 4b, the correlation is not 1:1 because of the $f$-$c$ hybridization. From results in Fig. 4, we conclude that spin-orbit coupling is a key factor in determining the maximum value of $T_c$ in the CeTX$_3$ family and that a narrow range of cell volumes, $i.e.$, hybridization, is necessary for inducing a superconducting state.

In summary, we have discovered superconductivity in heavy-fermion $CeRhGe_3$, the last non-superconducting member of the $CeTX_3$ family. This discovery has provides a unique opportunity to establish systematic correlations between superconductivity and materials properties within this family. It is not possible from our studies to make definitive statements about the superconducting mechanism, but it is clear that magnetic as well as valence fluctuations, and the spin-orbit coupling play an important role in pairing electrons. More experiments should be carried out to determine the evolution of the relative admixture of spin-singlet and spin-triplet pairing across the series, and to search for direct evidence of valence and magnetic fluctuations in the vicinity of $V_{crit}$.

Acknowledgements

The work in China was supported by the National Key Research and Development Program of China (Grant No. 2017YFA0303103, 2016YFA0300300, and 2017YFA0302900), the NSF of China (Grants No. 91321207, No. 11427805, No. 11404384, No. U1532267, No. 11604376, No. 11522435), the Strategic Priority Research Program (B) of the Chinese Academy of Sciences (Grant No. XDB07020300). Work at Los Alamos National Laboratory was performed under the auspices of the U.S. DOE, Office of Basic Energy Sciences, Division of Materials Sciences and Engineering.

† Correspondence and requests for materials should be addressed to L.S (llsun@iphy.ac.cn) and J.T (jdt@lanl.gov).

* These authors contributed equally to this work.

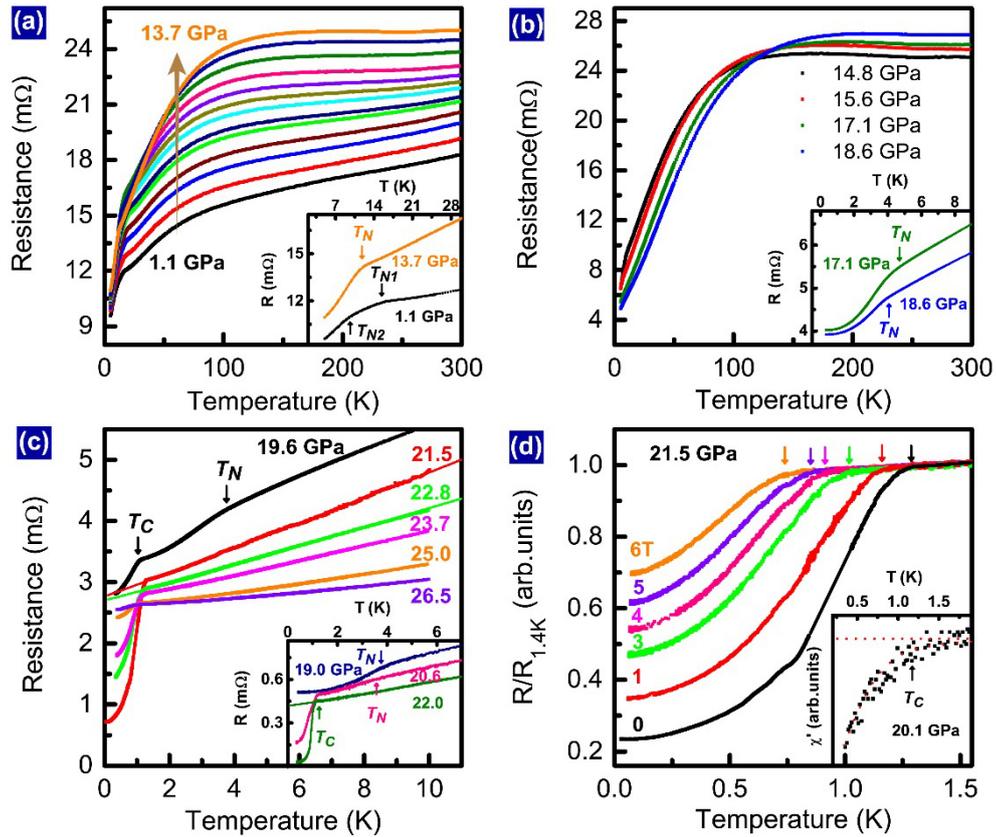

Figure 1 High-pressure electrical resistance and *ac* susceptibility of CeRhGe$_3$. (a) The temperature dependence of resistance at pressures from 1.1 to 13.7 GPa. The inset displays evidence for two antiferromagnetic transition temperatures, $T_{N1}$ and $T_{N2}$ (as indicated by arrows), for *P*=1.1 GPa that merge together at 13.7 GPa (as indicated by orange arrow). The second derivative of resistance with respect to temperature, plotted in Fig. S1, illustrates these transition most clearly. (b) The temperature dependence of resistance for pressures ranging from 14.8 GPa to 18.6 GPa. The inset

is an enlarged view of the resistance in the lower temperature regime at 17.1 GPa and 18.6 GPa, showing only one AFM transition at these two pressures. (c) The temperature dependence of resistance obtained in the pressure range 19.6 - 26.5 GPa. As discussed in the text, the pronounced drop in resistance at low temperatures reflects a pressure-induced superconducting transition. The inset displays the resistance as a function of temperature measured in another sample where the resistance drop at $T_c$ is more pronounced. The solid lines show the results of linear-$T$ fit. (d) The temperature dependence of the normalized resistance measured at 21.5 GPa under different magnetic fields. The arrows indicate the onset temperatures of the superconducting transitions. The inset shows the real part of the *ac* susceptibility measured at 20.1 GPa.

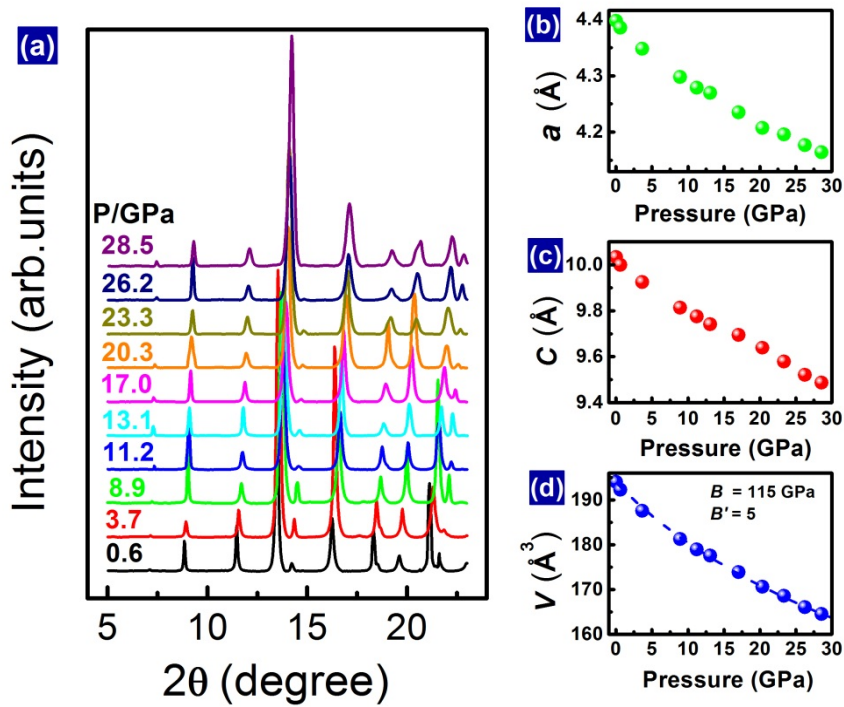

Figure 2 Structure of CeRhGe$_3$ under pressure. (a) X-ray diffraction patterns of CeRhGe$_3$ collected at different pressures, showing no structure transition up to 28.5

GPa. (b-d) Pressure dependence of lattice parameters *a, c* and unit cell volume (*V*). The dashed line in (d) is a fitting curve to the data by the Murnaghan equation of state that yields a bulk modulus $B$= 115 GPa and its pressure derivative $B'$=5.

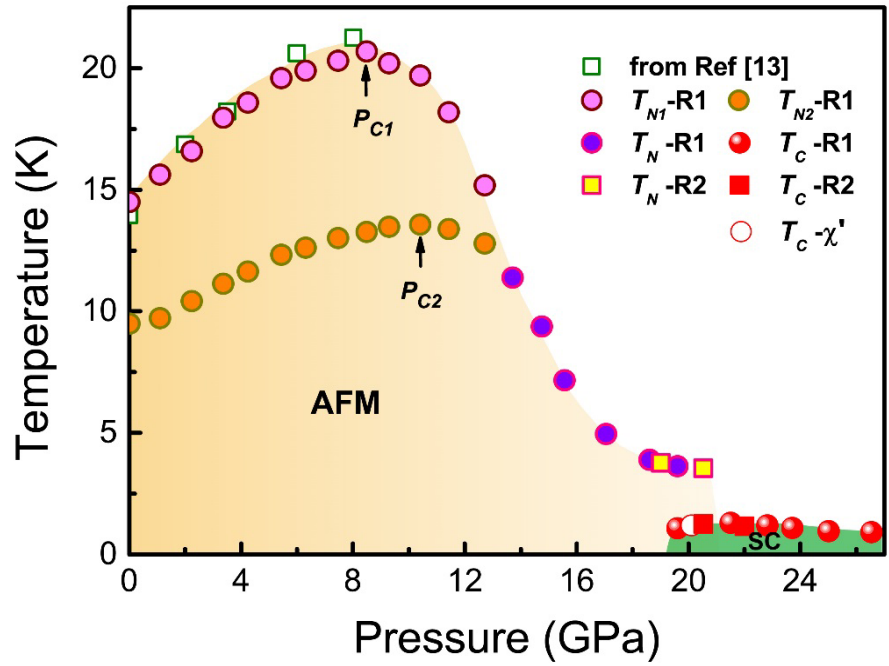

Figure 3 Temperature versus pressure phase diagram for $CeRhGe_3$. The solid circles in magenta and orange represent $T_{N1}$ and $T_{N2}$ determined by our high-pressure resistance measurements. $P_{C1}$ and $P_{C2}$ are the critical pressures where the $T_{N1}$ and $T_{N2}$ show a maximum value. The violet circles and yellow squares stand for the merged antiferromagnetic transition temperature ($T_N$) above 13.7 GPa. The solid red circles, squares and open circles denote the superconducting transition temperatures ($T_c$) determined by high-pressure resistance and *ac* susceptibility measurements. The olive squares are $T_{N1}$ transition temperatures adopted from Ref. [15].

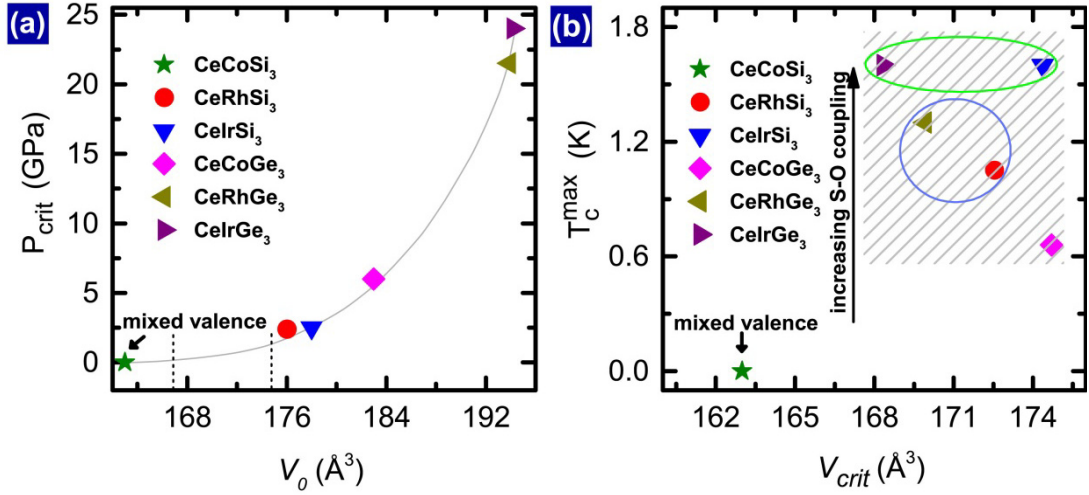

Figure 4 Correlations between the CeTX$_3$ cell volume and the maximum superconducting transition $T_c^{max}$. (a) Critical pressure $P_{crit}$ where $T_c$ reaches its maximum as a function of ambient-pressure unit cell volume $V_0$. The volume between the two vertical dashed lines along the abscissa corresponds to the range of volumes marked by the hashed box in (b). (b) Plot of maximum $T_c$ versus the critical unit-cell volume $V_{crit}$ where a maximum $T_c$ is observed. From the pressure dependence of the critical volume for CeRhGe$_3$ and the Murnaghan equation of state, we find the bulk modulus B and its derivative B' at zero pressure to be $B$=115 GPa and $B'$= 5. $V_{crit}$ for other family members are estimated using the same equation of state.